\documentclass[floats,floatfix,showpacs,amssymb,prd,twocolumn,superscriptaddress,nofootinbib,reprint]{revtex4-2}

\usepackage{amssymb,amsmath,verbatim,mathtools,needspace,enumitem,etoolbox,graphicx,physics,microtype,afterpage,bigints,gensymb,tabularx,xspace,bbm}

\usepackage{orcidlink} 
\usepackage[capitalise]{cleveref}
\usepackage{makecell}

\newcommand{\Yorsh}{\emph{Yorsh}}
\newcommand{\Sangria}{\emph{Sangria}}
\newcommand{\ein}{e_\mathrm{in}}
\newcommand{\fin}{f_\mathrm{in}}
\newcommand{\Mchirp}{\mathcal{M}_c}
\newcommand{\smBBH}{\textcolor{black}{SOBHB}}
\newcommand{\smBBHs}{\textcolor{black}{SOBHBs}}

\newcommand{\Tobs}{T_\mathrm{obs}}
\newcommand{\Tseg}{T_\mathrm{seg}}
\newcommand{\Nseg}{N_\mathrm{seg}}
\newcommand{\Ttr}{\Delta T}
\newcommand{\Ntr}{M_\mathrm{tr}}

\newcommand{\bham}{Institute for Gravitational Wave Astronomy \& School of Physics and Astronomy, University of Birmingham, Edgbaston, Birmingham B15 2TT, UK}

\newcommand{\AEI}{Max Planck Institute for Gravitational Physics (Albert Einstein Institute), Am Mühlenberg 1, 14476 Potsdam, Germany}

\newcommand{\cca}{Center for Computational Astrophysics, Flatiron Institute, 162 5th Avenue, New York, NY 10010}

\begin{document}

\title{
Global time-frequency search for stellar-mass binary black holes in LISA
}

\author{Diganta Bandopadhyay
\orcidlink{0000-0003-0975-5613}$^{\S}$}
\email{diganta.bandopadhyay@aei.mpg.de}
\affiliation{\AEI}
\affiliation{\bham}

\author{Christian E. A. Chapman-Bird
\orcidlink{0000-0002-2728-9612}$^{\S}$}
\email{c.chapman-bird@bham.ac.uk}
\affiliation{\bham}

\author{Alberto Vecchio
\orcidlink{0000-0002-6254-1617}}
\email{a.vecchio@bham.ac.uk}

\affiliation{\bham}
\affiliation{\cca}

\date{\today}

\begin{abstract}

    We present a complete pipeline for detecting and identifying gravitational waves (GWs) produced by the inspiral of stellar-mass binary black holes in data from the Laser Interferometer Space Antenna (LISA).
    The analysis framework relies on an efficient time-frequency implementation of an adaptive semi-coherent detection statistic, which we show to be robust against non-stationary noise and the presence of gaps of varying duration and cadence. 
    The search is able to detect signals with coherent signal-to-noise ratios as low as $\approx 11-14$ over the full parameter space of spin-aligned binaries (using non-spinning templates) with orbital eccentricity $\le 0.01$ when 
    deployed on the 2-year-long LISA Data Challenge \Yorsh{}. The search can be run within a day using $\approx 40$ GPUs. The techniques presented here have wider applications in GW astronomy, in particular the search for extreme-mass-ratio inspirals in LISA data.

\end{abstract}

\maketitle
\begingroup
\renewcommand\thefootnote{}
\footnotetext{$^{\S}$~These authors contributed equally to this work.}
\endgroup

\section{Introduction}
The challenge of performing a global (\textit{i.e.} full parameter space) search for gravitational waves (GWs) from binary systems is ubiquitous in GW astronomy. 
This affects ground-based observations of neutron stars and black holes (see \textit{e.g.} Ref.~\cite{LIGOScientific:2025yae} and references therein) as the LIGO~\cite{aLIGO}, Virgo~\cite{aVirgo} and KAGRA~\cite{KAGRA} detectors (and more ambitious future generation observatories~\cite{ET,CE}) improve their low-frequency sensitivity and the search domain is extended to sub-solar mass objects \cite{LVK_subsolar:2023}, orbital eccentricity \cite{2024ApJ...973..132A} and arbitrary black hole spins \cite{2016PhRvD..94b4012H}. 
It is also an unsolved problem for the Laser Interferometer Space Antenna (LISA)~\cite{2017arXiv170200786A, 2024arXiv240207571C} in the context of identifying  stellar-origin ($\sim 1 - 100\,\mathrm{M}_\odot$) black hole binaries (\smBBHs) $\approx 1-100\,\mathrm{yr}$ from coalescence, which would provide unique insight into their astrophysical formation pathways~\cite{Sesana:2016, Gerosa:2019, Moore:2019, Klein:2022,2024arXiv240207571C}, and extreme-mass-ratio inspirals (EMRIs) of neutron stars and stellar-mass black holes orbiting massive black holes ($\sim 10^5 - 10^7\,M_\odot$), which represent new probes of the formation and evolution of massive black holes at the centre of galaxies, their environment~\cite{LISA:2022yao},  and unique laboratories for tests of general relativity~\cite{LISA:2022kgy}. 

Currently, a computationally viable search strategy is unknown for \smBBHs{} and EMRIs. The parameter space volume covered by the likelihood function of typical signals is a minuscule fraction of the astrophysical prior, from $\sim \mathcal{O}(10^{-32})$ for \smBBHs{} to $\mathcal{O}(10^{-46})$ for EMRIs (see Fig.~1 in Ref.~\cite{Moore:2019}). Their complex signals, originating from spinning black holes in binaries with eccentric orbits, have power distributed across (potentially) many modes -- $\sim 10^3$, in the case of EMRIs -- and must be integrated over $\sim 10^5$ wave cycles to accumulate sufficient signal-to-noise ratio (SNR) to be detectable. 
These factors make standard match-filtering searches unfeasible~\cite{Moore:2019,Gair:2004iv, Chua:2021aah}. 
Alternative methods, including excess-power~\cite{Strub:2025dfs,Gair:2004iv,Gair:2006nj,Wen:2005xn}, machine-learning~\cite{Zhang:2022xuq,Cole:2025sqo} or phenomenological fits~\cite{Ye:2023lok,Wang:2012xh} are being explored, but a viable solution has yet to be found.

The problem of identifying and characterizing \smBBHs{} in LISA data is under active research~\cite{Buscicchio:2021, Digman:2023, Bandopadhyay:2023, Bandopadhyay:2024,Bandopadhyay:2025,Wang:2025,Wang:2024,Wu:2025zhc}. 
These sources offer key insights into both astrophysics and fundamental physics~\cite{Buscicchio:2021,Toubiana:2020vtf,Gupta:2020lxa}, complementing the late-inspiral observations made by ground-based detectors~\cite{2016PhRvL.116f1102A, 2016PhRvX...6d1015A, 2019PhRvX...9c1040A, 2020PhRvL.125j1102A, 2021PhRvX..11b1053A, 2023PhRvX..13d1039A, 2025arXiv250708219T, LIGOScientific:2025slb, KAGRA:2025oiz}. Their weak-field nature and comparable mass ratios further make them an ideal test-bed for developing techniques that may ultimately be applied to the EMRI search. 
In addition, \smBBHs{} are expected to be rare~\cite{Buscicchio:2025}, which pushes demands on search strategies to be nearly optimal in terms of sensitivity. Furthermore, their radiation leaves the LISA band $\sim \mathrm{weeks}$ before the black holes merge, which places severe time-constraints on any global search implementation to allow telescopes to point \textit{in advance} and stare at a merger unfolding~\cite{deMink:2017,Klein:2022, Yi:2019}, while being also observed in GWs by ground-based detectors.

In this work, we present an end-to-end global search method based on an adaptive semi-coherent statistic~\cite{Bandopadhyay:2024,Bandopadhyay:2025}. 
We demonstrate that this method is able to detect \smBBH{} signals in year-long LISA (mock) datasets down to $\mathrm{SNR} \approx 11$ over the full astrophysical parameter space of mildly-eccentric, spin-aligned \smBBHs{} (using non-spinning templates). 
At the core of our search is a hardware-accelerated time-frequency domain waveform representation -- a natural basis for broad-band, slowly-chirping signals~\cite{Digman:2023,Tenorio:2025} -- that reduces  computational cost by $\sim 10^3$ with respect to frequency-domain methods.   
As shown in \Cref{tab:search-comparisons}, this pipeline searches the entirety of the two-year LISA Data Challenge (LDC) \Yorsh{} data set in $\lesssim 12$ days on 4 NVIDIA A100 graphics processing units (GPUs). 
Due to its computational speed-up of $\sim250\times$ over the previous state-of-the-art~\cite{Bandopadhyay:2024,Bandopadhyay:2025} and it being massively parallelisable, it can achieve sub$-1\,\mathrm{day}$ latency on $\approx 40$ GPUs. 
We also demonstrate that our approach naturally deals with (and is robust to) data gaps and non-stationary noise sources, which introduce additional challenges in searching for long-lived signals. 
It is therefore readily applicable to realistic LISA data.

This work presents a practical solution to the long-standing challenge of \smBBH{} search, establishes a foundation for tackling the more complex problem of EMRI detection, and several of its underpinning techniques have broader applicability to LIGO-Virgo-KAGRA searches for compact binaries over increasingly wider parameter space. 
In this study, we use geometric units ($c=G=1$).

A high level overview of the end-to-end search pipeline described here is illustrated in Fig.~2 of Ref.~\cite{Bandopadhyay:2024}.

\begin{table}
    \caption{
    Computational costs -- timed on NVIDIA A100 GPUs -- of the proposed pipeline. The detection statistic $\Upsilon_{\Nseg}$ is defined in \cref{eq:Upsilon} and (\ref{eq:inner-tf}). Each search tile is assigned to (and timed on) a single GPU. The search of the \Yorsh{} dataset presented here used 4 GPUs. 
    We also provide comparison with the pipeline described in Ref.~\cite{Bandopadhyay:2025} (BM25). 
    Only the cost of evaluating $\Upsilon_{\Nseg}$ is directly comparable; the BM25 total wall-time is therefore extrapolated assuming 4 GPUs were used.
    }
\begin{ruledtabular}
\begin{tabular}{|l|cc||}
    {} & This work & BM25 \\
    \hline \hline 
    Cost per $\Upsilon$ evaluation $[\mu\mathrm{s}]$ & $1-10$ & $10^3-10^4$ \\
    No. of $\Upsilon$ evaluations per search tile & $10^8$ &  \\
    Wall-time per tile [h] & $0.1-0.5$ &  \\
    Total number of search tiles & $2680$ &  \\

    Full search wall-time on 4 GPUs [d] & $11$ & $\sim$ \textit{3000} \\
\end{tabular}
\end{ruledtabular}
\label{tab:search-comparisons}
\end{table}

%
%

%
%

\section{Methods}

In this section, we introduce the detection statistic, its efficient time-frequency representation and the GPU-optimised acceleration of its implementation that we adopt to build an end-to-end search for \smBBHs{}.

Our approach exploits a hierarchy of timescales, $\Tobs \ge \Tseg \ge \Ttr$, that we introduce here:
(i) $\Tobs$ is the \textit{total} duration of the data set that for LISA is several years (the nominal mission lifetime is $4.5\,\mathrm{yr}$ with possible extension to $10\,\mathrm{yr}$~\cite{2024arXiv240207571C}); we set the (arbitrary) reference time $t_0 = 0$ at the start of observation, so that the time of coalescence of a \smBBH, $t_c$, corresponds to the their remaining lifetime; (ii) $\Tseg$ is the duration of the non-overlapping \textit{time segments} into which we divide the data set, and defines the coherence length of our analysis; the total number of segments (which varies as the search progresses) is $\Nseg \equiv \tau/ \Tseg = 100\,(\tau/10^8\,\mathrm{s})\, (10^6\,\mathrm{s}/\Tseg)$ for $\tau=\mathrm{min}(\Tobs,t_c)$; (iii) $\Ttr$ is the length of the data \textit{tranche} over which an efficient time-frequency approximation of the inner product between the data and the signal's templates is computed, and is set in advance ensuring that second-order contributions to the Taylor expansion of the signal phase $\approx 0.2\,\left(\Mchirp/{30\,M_\odot}\right)^{10/3}\,\left({f}/{10\,\mathrm{mHz}}\right)^{19/3}\,\left({\Ttr}/{10^4\,\mathrm{sec}}\right)^{3}$ can be neglected; here $\Mchirp$ is the binary's chirp mass and the total number of tranches in a segment is $\Ntr \equiv \Tseg/\Ttr$.

\subsection{Time-frequency search statistic}\label{sec:time-frequency-statistic}

In this section we detail the procedure for constructing a time-frequency formulation of the detection statistic. Crucially, this relies on a sparse representation of the waveform which is derived here. We begin by defining the statistic in the frequency domain and then describe its efficient computation in the time-frequency domain.

Given two real time series $a(t)$ and $b(t)$, with Fourier transform $\tilde a(f)$ and $\tilde b(f)$, respectively, we define the usual noise-weighted inner-product
\begin{equation}
    \langle a \mid b \rangle = 4\, \mathrm{Re}  \int_0^\infty \frac{\tilde{a}^*(f)\,\tilde{b}(f)}{S(f)} \mathrm{d}f\,,
    \label{eq:inner-cont}
\end{equation}
where $S(f)$ is the one-sided noise power spectral density (PSD). We also introduce
\begin{equation}
    (a \mid b ) = \int_0^\infty \frac{\tilde{a}^*(f)\,\tilde{b}(f)}{S(f)} \mathrm{d}f\,,
    \label{eq:inner-complex-cont}
\end{equation}
which, with slight abuse of terminology, we call the complex inner product, such that $\langle a | b \rangle = 4 \mathrm{Re} \left\{ (a | b ) \right\}$\,.

We make the standard assumption that the three Time-Delay Interferometry (TDI) channels, labelled by $\alpha$, are pseudo noise-orthogonal~\cite{2021LRR....24....1T} (although this can be relaxed at minor additional cost). 
From Eq.~(\ref{eq:inner-cont}), the coherent inner product between the data, $d = \{d_\alpha\}$, and a GW signal template characterised by parameters ${\mathbf \theta}$, $y({\mathbf \theta}) = \{y_\alpha({\mathbf \theta})\}$, is
\begin{multline}
    \langle d \mid y(\mathbf{\theta}) \rangle = \sum_{\alpha = 1}^3 \langle d_\alpha \mid y_\alpha(\mathbf{\theta}) \rangle \\
    = \sum_{\alpha = 1}^3 4\, \Re \left\{ \int_0^\infty \frac{\tilde{d}_\alpha^*(f)\,\tilde y_\alpha(\theta; f)}{S_\alpha(f)} \mathrm{d}f\right\}\,,
    \label{eq:inner-all-cont}
\end{multline}
where the noise weighting function for $\langle d_\alpha \mid y_\alpha(\mathbf{\theta}) \rangle$ is the one-sided noise PSD in the relevant TDI channel, $S_\alpha(f)$.

One of the parameters that define the signal in the data is an overall constant phase, $\phi$, at some (arbitrary) reference time. Writing $\theta = \{\xi, \phi\}$, such that $\xi$ is the vector of parameters \textit{excluding} an overall constant phase, and substituting Eqs.(\ref{eq:inner-cont}) and~(\ref{eq:inner-complex-cont}) into Eq.~(\ref{eq:inner-all-cont}), one can maximise the inner-product over this phase as
\begin{align}
    {\cal O}(\xi) & = \max_\phi \langle d \mid y(\mathbf{\xi}) e^{i\phi} \rangle 
    \nonumber \\
    & = \max_\phi \sum_{\alpha = 1}^3 \langle d_\alpha \mid y_\alpha(\mathbf{\xi}) e^{i\phi} \rangle  
    \nonumber \\
    & = 4 \left| \sum_{\alpha = 1}^3 (d_\alpha \mid y_\alpha(\mathbf{\xi})) \right|.
    \label{eq:overlap}
\end{align}

The quantity ${\cal O}(\xi)$ is also known as the \textit{overlap} of the template with the data, and the last line in Eq.~(\ref{eq:overlap}) shows that phase maximisation can be performed analytically and efficiently in the Fourier domain.

In this work, we employ the semi-coherent statistic introduced in~\cite{Bandopadhyay:2023, Bandopadhyay:2024,Chua:2017ujo}. 
Segmenting the data such that the $n-$th segment covers the time interval $[t_n, t_{n+1}]$, where $t_n = n\,\Tseg$, the detection statistic is: 
\begin{align}
    \Upsilon_{\Nseg}(\xi) & = \sum_{n = 0}^{\Nseg - 1} \frac{\left[\max_{\phi_n} \langle d_{n} \mid y_{n}(\mathbf{\xi}) e^{i\phi_n} \rangle\right]^2}{\langle y_{n}(\mathbf{\xi}) \mid y_{n}(\mathbf{\xi}) \rangle}\,,
    \label{eq:Upsilon}
\end{align}
where we have defined the inner-product over the segment (for all TDI channels) as:
\begin{equation}
    \left\langle d_n \mid y_n e^{i\phi_n}\right\rangle = 4\, \mathrm{Re} \left\{e^{i\phi_n} \sum\limits_{\alpha=1}^{3} 
    \sum\limits_{k=0}^{N_f-1}\frac{\tilde{d}_{\alpha,n}^*[k]\,\tilde{y}_{\alpha,n}[k]}{S_{\alpha,n}[k]}\Delta f \right\}
    \,,
    \label{eq:inner}
\end{equation}
and analogously for $\langle y_n \mid y_n\rangle$. Here $\tilde{x}$ denotes the Fourier transform of a time series $x$, $\Delta f = 1/\Tseg$, $N_f = \Tseg/\Delta t$, $\Delta t$ is the sampling cadence, and the index $k$ corresponds to the discrete frequency $f_k = k\Delta f$. $S_{\alpha, n}$ is the one-sided noise power spectral density (PSD) in the segment. 

In the definition of the detection statistic we have explicitly split the vector of unknown parameters that characterise a signal into $\theta = \{\xi, \phi\}$, where $\xi$ is the vector of parameters \textit{excluding} a constant phase offset $\phi$. Note that we are also maximizing over an amplitude parameter implicitly within each segment in the definition of ~\cref{eq:Upsilon}. For a given $\Nseg$, the unknown parameters are $(\xi, \{\phi_n, A_n; n=1,\dots, \Nseg\})$, and therefore the effective dimensionality of the parameter space is $\dim(\xi) + 2\Nseg$, with maximisation over $2\Nseg$ parameters. 
As the search progresses, $\Nseg$ is gradually reduced; the final stage, $\Nseg = 1$, is fully coherent and the search statistic is $\Upsilon_{1} \equiv \mathrm{max}_{\phi}(\rho^2)$, where $\rho$ is the matched filter SNR, this hierarchical process reduces the false-alarm probability of the search~\cite{Bandopadhyay:2024}.

The key to our search is efficiently computing~\cref{eq:inner}, which we achieve with a hardware-accelerated framework in the time-frequency domain under the assumption of a slowly-evolving GW frequency. 
We divide each segment in tranches, so that the $m-$th tranche covers the time interval $[t_{nm}, t_{n\,m+1}]$ with $t_{nm} = n\Tseg + m \Ttr$. 
We can now write the inner product Eq.~(\ref{eq:inner}) as:
\begin{eqnarray}
    \left\langle d_n \mid y_n e^{i\phi_n}\right\rangle & = & 4\, \mathrm{Re} 
    \left\{e^{i\phi_n}\sum\limits_{\alpha=1}^{3} 
    \sum\limits_{m=0}^{\Ntr-1}
    \mathcal{T}_{\alpha,nm} \right. \nonumber \\
    && \times \left.\sum\limits_{k=1}^{N_\mathcal{F}}
    \frac{\tilde{d}_{\alpha, nm}^*[k]\, \tilde{h}_{\alpha, nm}[k]}{S_{\alpha, nm}[k]}\delta f\right\}\,,
    \label{eq:inner-tf}
\end{eqnarray}
with $\delta f = 1 / \Delta T$. Equation~\ref{eq:inner-tf} is significantly more efficient to compute than Eq.~(\ref{eq:inner}) because (i) the sum over frequency is compact and can be truncated to $N_\mathcal{F} \ll \Ttr/\Delta t$ bins, centred on $f_{nm} \equiv f(t_{nm})$; (ii) the frequency-domain waveform $\tilde{h}_{nm}(f)$ is expressed analytically in terms of Fresnel integrals; (iii) the TDI transfer functions $\mathcal{T}_{\alpha,nm} \equiv \mathcal{T}_\alpha(f_{nm}) $ are complex multiplicative factors, computed once per tranche for all frequency bins. 
As shown in \cref{tab:search-comparisons}, under assumptions suitable for \smBBHs{} (and GPU acceleration), this approach is orders of magnitude more efficient than computation of~\cref{eq:inner} in the frequency domain, rendering a global \smBBH{} search feasible.

The detection statistic, Eq.~(\ref{eq:Upsilon}), is the sum of (suitably normalised) overlaps over the time segments into which we partition the data set. 
We divide the data set of duration $\Tobs$ in non-overlapping time segments of length $\Tseg$, where the total number of segments is $\Nseg = \Tobs/\Tseg$.  Using Eq.~(\ref{eq:overlap}), the detection statistic~(\ref{eq:Upsilon}) can be written as  
\begin{align}
    \Upsilon_{\Nseg}(\xi) & = \sum_{n = 0}^{\Nseg - 1} 
    \frac{{\cal O}_n^2(\xi)}{\sum_{\alpha = 1}^3 \langle y_{\alpha,n}(\mathbf{\xi}) \mid y_{\alpha,n}(\mathbf{\xi}) \rangle}\,,
    \label{eq:Upsilon-seg}
\end{align}    
where we have defined the overlap over the segment as
\begin{align}
    {\cal O}_n(\xi) & = \max_{\phi_n} \sum_{\alpha = 1}^3 \langle d_{\alpha,n} \mid y_{\alpha,n}(\mathbf{\xi}) e^{i\phi_n} \rangle\,,
    \nonumber \\
    & = 4 \left| \sum_{\alpha = 1}^3 (d_{\alpha,n} \mid y_{\alpha,n}(\mathbf{\xi})) \right|\,.
    \label{eq:overlap-n}
\end{align}    
In the previous equation, $\langle d_{\alpha,n} \mid y_{\alpha,n}(\mathbf{\xi}) e^{i\phi_n} \rangle$ is given by Eq.~(\ref{eq:inner}), and in analogy with \cref{eq:inner-complex-cont} we have defined the segment-level complex inner product
\begin{align}
    \left(d_{\alpha,n} \mid y_{\alpha,n}(\xi) \right) & = \int_0^\infty \frac{\tilde{d}_{\alpha,n}^*(f)\,\tilde y_{\alpha,n}(\xi; f)}{S_{\alpha, n}(f)} \mathrm{d}f \,,
    \nonumber\\
     & = \sum\limits_{k=0}^{N_f-1}\frac{\tilde{d}_{\alpha,n}^*[k]\,\tilde{y}_{\alpha,n}(\xi)[k]}{S_{\alpha,n}[k]}\Delta f   \,, 
     \label{eq:inner-segment}
\end{align}
to emphasise the process by which one maximises over the $\Nseg$ phase parameters.
In Eqs.~(\ref{eq:Upsilon-seg}), (\ref{eq:overlap-n}) and~(\ref{eq:inner-segment}) we have defined a time-series $x_{\alpha, n}$ as the output of the TDI channel $\alpha$ over the time segment $n$. The data are discretely sampled at times $t_j = j\,\Delta t$, where $\Delta t$ is the sampling cadence. We denote the discrete time series as $x_{\alpha, n}[j] = x(n\,\Tseg + j\Delta t)$. The discrete Fourier components of $x_{\alpha, n}$ at frequency $f_k = k \Delta f$ are denoted by $\tilde x_{\alpha, n}[k]$, where $\Delta f = 1/\Tseg$ and $N_f = \Tseg/\Delta t$; $S_{\alpha, n}[k]$ is the one-sided noise spectral density of the data at frequency $f_k$.

\subsection{Time-frequency waveform representation}

It is essential to compute Eq.~(\ref{eq:inner-segment}) efficiently, which we do by exploiting the slow phase evolution of GWs from \smBBHs{} over $\Tobs$.
Assuming one waveform mode is dominant, the complex GW strain can be written as
\begin{equation}
    h(t) = \mathcal{A}(t) e^{i\varphi(t)}.
    \label{eq:ht}
\end{equation}
We divide each segment into tranches of duration $\Ttr$. 
For the $n-$th segment, the $m-$th tranche covers the time interval $[t_{nm}, t_{n\,m+1}]$, where $t_{nm} = n\Tseg + m \Ttr$. 
Within each tranche, we Taylor-expand the GW phase to second order:
\begin{align}
    \varphi(t) & = \varphi_{nm} + 2\pi f_{nm} (t - t_{nm}) + \pi \dot{f}_{nm} (t - t_{nm})^2
    \nonumber \\
    & \quad 
    + {\cal O}((t - t_{nm})^3)\,,
\end{align}
where
\begin{align}
    \varphi_{nm} & \equiv \varphi(t_{nm})\,, \\
    f_{nm} & \equiv f(t_{nm}) = \frac{1}{2\pi} \left[\frac{d\varphi}{dt}\right]_{t = t_{nm}}\,, 
\end{align}
and
\begin{equation}
    \dot{f}_{nm} \equiv \frac{1}{2\pi} \left[\frac{d^2\varphi}{d^2t}\right]_{t = t_{nm}}\,.
\end{equation}
The Fourier transform of \cref{eq:ht} over the tranche can now be approximated as
\begin{equation}
    \tilde{h}_{nm}(f) = \frac{\mathcal{A}_{nm}}{\sqrt{2\dot{f}_{nm}}}e^{i(\varphi_{nm}- \pi \dot{f}_{nm}\zeta_{nm}^2(f))}\mathcal{F}_{nm}(f)\,,
    \label{eq:fresnel-ft}
\end{equation}
where
\begin{align}
    \zeta_{nm}(f) & = \frac{f_{nm} - f}{\dot{f}_{nm}}\,,\\
    \mathcal{A}_{nm} & = \mathcal{A}(t_{nm})\,.
\end{align}
In Eq.~(\ref{eq:fresnel-ft}) we have defined the Fresnel kernel function
\begin{equation}
    \mathcal{F}_{nm}(f) = \Delta C_{nm}(f) + i \Delta S_{nm}(f)\,,
    \label{eq:kernel-function}
\end{equation}
with
\begin{align}
    \Delta C_{nm}(f) & = C[v_{nm}(t_{n\,m+1}, f)] - C[v_{nm}(t_{nm}, f)]\,,
    \\
    \Delta S_{nm}(f) & = S[v_{nm}(t_{n\,m+1}, f)] - S[v_{nm}(t_{nm}, f)],
    \\
    v_{nm}(t,f) & = \sqrt{2\dot{f}_{nm}} \,\left[(t - t_{nm}) + \zeta_{nm}(f)\right]\,,
\end{align}
and $S(z) = \int_0^z \sin(\pi x^2/2) dx$ and $C(z) = \int_0^z \cos(\pi x^2/2) dx$ are the Fresnel integrals~\cite{Abramowitz1964}. 

A crucial aspect for the efficient computation of \cref{eq:inner-segment} is the behaviour of the kernel function~(\ref{eq:kernel-function}) as a function of frequency: for the values of $\dot{f}_{nm}$ typical of \smBBHs{} in the relevant frequency band $\sim 1 - 100\,\mathrm{mHz}$, the waveform and thus the Fresnel-kernel in \cref{eq:kernel-function} (and hence \cref{eq:fresnel-ft}) peak sharply around $f_{nm}$ and fall off rapidly in frequency (see extended discussion on the compactness of this kernel in Appendix~\ref{app:compactness}).
Additionally, as $v_{nm}(t, f) \sim \delta f / \sqrt{\dot{f}}$ is typically large ($\sqrt{\dot{f}} \ll \delta f$ for much of the inspiral), we specifically employ asymptotic forms of the Fresnel integrals to further reduce computational costs~\cite{NIST:DLMF}.
For argument $|z| > 6$, these take the form
\begin{equation}
    S(z) = \frac{1}{2} - \frac{1}{\pi z}\cos\left(\frac{\pi z^2}{2}\right)
\end{equation}
and
\begin{equation}
    C(z) = \frac{1}{2} + \frac{1}{\pi z}\sin\left(\frac{\pi z^2}{2}\right),
\end{equation}
agreeing to better than one part in $10^4$ for all $|z| > 6$.
Note that $S(-z) = -S(z)$ and $C(-z) = - C(z)$.
For $|z| \leq 6$, we use the modified forms
\begin{equation}
    S(z) = \frac{1}{2} - f(z) \cos\left(\frac{\pi z^2}{2}\right) - g(z)\sin\left(\frac{\pi z^2}{2}\right)
\end{equation}
and
\begin{equation}
    C(z) = \frac{1}{2} + f(z) \sin\left(\frac{\pi z^2}{2}\right) - g(z)\cos\left(\frac{\pi z^2}{2}\right),
\end{equation}
with
\begin{equation}
    f(z) = \frac{1 + 0.926z}{2 + 1.792z + 3.104z^2}
\end{equation}
and
\begin{equation}
    g(z) = \frac{1}{2 + 4.142z + 3.492z^2+6.670z^3}.
\end{equation}
The polynomial coefficients are sourced from Ref.~\cite{MCCORMICK2002261}.
This extends the agreement to one part in $10^4$ over the remainder of the domain in $z$.

Finally, as the last step in constructing the signal model, we write the TDI observable as 
\begin{equation}
    y_{\alpha,nm}(f)= \mathcal{T}_\alpha(f_{nm})\,\tilde{h}_{nm}(f).
    \label{eq:tdi2}
\end{equation}
where $\mathcal{T}_{\alpha}$ are TDI transfer functions, and we have explicitly re-introduced the index $\alpha$ to label the different TDI variables. Note that the transfer function is only computed for the `central' frequency $f_{nm}$ within each tranche and applied to the whole waveform $\tilde{h}(f)$, making use of the relative compactness of $\tilde{h}(f)$ compared to variations in $\mathcal{T}_\alpha(f)$. The waveform is sharply peaked around its central frequency over hour-long timescales, meanwhile the instrument response function varies relatively smoothly and has support at a wide range of frequencies, see for example Fig.~8 in Ref.~\cite{Marsat:2018}.

\Cref{eq:inner-segment} can now be computed by combining the analogous quantities computed at the tranche level:
\begin{equation}
    (d_{\alpha, n} | y_{\alpha, n}(\xi)) = \sum_{m = 0}^{\Ntr - 1} (d_{\alpha, nm} | y_{\alpha, nm}(\xi)))\,,
    \label{eq:inner-segment-tf}
\end{equation}
where
\begin{equation}
    (d_{\alpha, nm} | y_{\alpha, nm}(\xi)) = \mathcal{T}_\alpha(f_{nm}; \xi)\sum\limits_{k=1}^{N_\mathcal{F}}\frac{\tilde{d}_{\alpha, nm}^*[k]\, \tilde{h}_{\alpha, nm}(\xi)[k]}{S_{\alpha, nm}[k]}\delta f\,,
    \label{eq:inner-tranche-tf}
\end{equation}
and $\Ntr = \Tseg/\Ttr$ is the number of tranches in each segment, $\delta f = 1/\Ttr$, and $N_\mathcal{F} \ne N_f$ is the number of frequency components over which the summation is performed. Analogously, Eq.~(\ref{eq:inner-tf}) can be written in the form:
\begin{eqnarray}
    \left\langle d_n \mid y_n(\xi) e^{i\phi_n}\right\rangle & = & 4\, \mathrm{Re} 
    \left\{e^{i\phi_n}\sum\limits_{\alpha=1}^{3} (d_{\alpha, n} | y_{\alpha, n}(\xi)) \right\} \nonumber \\
    & = & 4\, \mathrm{Re} 
    \left\{e^{i\phi_n} \right.
    \nonumber \\
    & & \times \left. 
    \sum\limits_{\alpha=1}^{3} \sum\limits_{m=0}^{\Ntr-1} 
    (d_{\alpha, nm} | y_{\alpha, nm}(\xi))\right\}\,.
\end{eqnarray}

\Cref{eq:inner-segment-tf} is significantly more efficient to compute than \cref{eq:inner-segment} for several reasons. 
The evaluation of \cref{eq:tdi2} does not require a (relatively) expensive numerical Fourier transform, but it is semi-analytical (via \cref{eq:fresnel-ft}) and the LISA transfer function enters just as a (complex) multiplicative factor. 
The data are Fourier transformed only once per tranche. Finally, the sum over the frequency components within the tranche inner-product~(\ref{eq:inner-tranche-tf}) is not done over the full frequency grid but on a much smaller number of frequency bins $N_\mathcal{F} \ll \Ttr/\Delta t$. 

In this work, we find that evaluating \cref{eq:fresnel-ft} over $N_\mathcal{F} = 21$ bins centred on the frequency $\lceil f / f_{nm}\rceil \delta f$ (where $\lceil \rceil$ is the ceiling operator, \textit{i.e.} the Fresnel kernel is centred on the nearest frequency bin above $f_{nm}$), instead of all $2160\, (\Ttr/6\,\mathrm{hr})\,(5\,\mathrm{s}/\Delta t)$ bins, is sufficient to recover $>99\%$ of the waveform power. 
This is a hundred-fold reduction in data volume with respect to the full frequency grid, greatly reducing the cost of \cref{eq:Upsilon-seg}.

\subsection{Hardware acceleration and computational cost}
We can further improve the computational efficiency of \cref{eq:inner-segment-tf}, and therefore reduce the overall cost of the evaluation of the detection statistic, by taking advantage of GPU hardware.
To do so, we must first identify what stage of the computation should be distributed among each block of GPU threads.
The architecture of modern GPUs typically requires specifying a size of $32^n$ threads per block (where $n\in \mathbb{N}$), so imposing parallelism at the level of the evaluation of ~\cref{eq:inner-tranche-tf} over Fourier bins within a tranche would significantly limit flexibility.
One could instead parallelise the computation with respect to the tranche index $m$, as multiple years of LISA data consists of thousands of these elements.
This choice is best when the desired application is to evaluate statistics for $\lesssim 10^2$ sets of waveform parameters in parallel, as it makes full use of the GPU thread-count, but incurs additional memory management and synchronisation costs as many thread blocks must collaborate to evaluate the statistic for each set of source parameters.

\begin{figure}[h]
    \centering
    \includegraphics[width=0.99\linewidth]{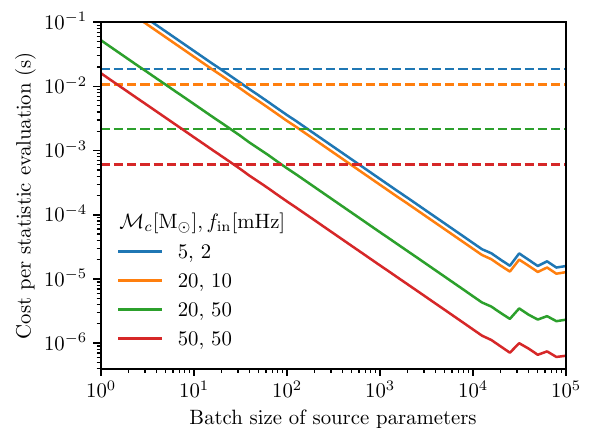}
    \caption{Wall-time per set of source parameters in the evaluation of \cref{eq:Upsilon-seg} as a function of batch size for $N_\mathcal{F} = 21$ (see~\cref{eq:inner-segment-tf}), $\Tobs=2\,\mathrm{yr}$ and $\Delta t=5\,\mathrm{s}$.
    The implementation on an NVIDIA A100 GPU (solid) outperforms the one on a $2.4\,$GHz Intel Xeon Platinum single-core CPU (dashed lines) starting from a batch size $\approx 30$ (single core). Wall-times are averaged uniformly with respect to mass ratio and eccentricity within the bounds targeted by our search.
    The oscillations in the wall-time for batches larger than $\sim 10^4$ are due to saturation of GPU resources.
    }
    \label{fig:timing-scaling}
\end{figure}

In our pipeline, we typically evaluate more than $10^5$ detection statistics at once to rapidly explore the \smBBH{} parameter space. 
This motivates parallelising \cref{eq:Upsilon-seg} at the highest level possible, devoting one GPU thread to each set of source parameters.
Detection statistics can then in principle be computed en-masse for the entire batch at each iteration of the optimisation algorithm, achieving extremely low wall-times per parameter set.
However, na\"ively attempting to compute $10^5$ waveforms in parallel would entail the construction of a matrix of waveforms of $(10^5 \times \Nseg \times \Ntr \times N_\mathcal{F}) \approx 10^{10}$ elements, which is expensive and too large for the global memory of most GPUs.

Rather than storing waveforms and then computing statistics as two separate operations, we address this limitation by storing values of $\Upsilon_{\Nseg}$ that are updated directly via the inner products $\langle d_{\alpha, n} | y_{\alpha, n}\rangle$ and $\langle y_{\alpha, n} | y_{\alpha, n}\rangle$, tranche by tranche, as the waveform is constructed.
In doing so, no waveform storage is necessary, and the large matrix of waveforms is replaced with a vector of $10^5$ statistic values (one per set of source parameters).
This level of memory management also necessitates the use of low-level \textsc{cuda} kernels, as opposed to the use of high-level frameworks such as \textsc{jax}.
Direct evaluation of $\Upsilon_{\Nseg}$ is an essential component of our methodology --- it minimises the number of memory allocations required, which ultimately bottlenecks GPU operations, and facilitates sufficiently large batch sizes for \cref{eq:Upsilon-seg} to effectively be computed in microseconds.
This is demonstrated in \cref{fig:timing-scaling}. We observed that speed-ups of a factor $\sim 10^3$ with respect to a single-core CPU are achieved when the batch size is larger than $\sim 10^4$.
The resulting algorithm is also extremely lightweight with respect to global memory usage, requiring $\lesssim 1\,\mathrm{GB}$ of memory even for large batch sizes; it is therefore well-suited to deployment on a large number of lower-grade GPUs, for which the memory available can be limited to a few $\mathrm{GB}$.

\begin{figure}[h]
    \centering
    \includegraphics[width=0.99\linewidth]{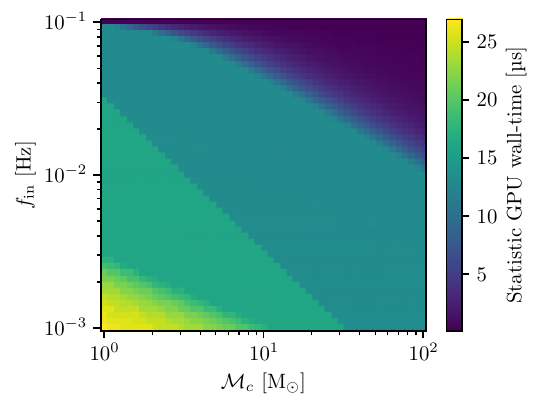}
    \caption{
    Computational wall-time per evaluation of \cref{eq:Upsilon-seg} in the $(\mathcal{M}_c, f_\mathrm{in})$ plane, averaged over the mass ratio and eccentricity ranges probed by our pipeline.
    The upper-right corner of the plot (where costs are lowest) corresponds to signals that exit the LISA band during the observational window.
    }
    \label{fig:timing-2d}
\end{figure}

In \cref{fig:timing-2d}, we show the GPU wall-time per statistic computed (for a batch size of $10^4$) in the $(\Mchirp, \fin$) plane explored by our search pipeline.
Performance varies from $\sim 1\,\mu\mathrm{s}$ for higher $(\mathcal{M}_c, \fin)$, to $\sim 25\,\mu\mathrm{s}$ for lower $(\Mchirp, \fin)$, over the parameter space.
The fuzzy boundary (due to averaging over other source parameters) in \cref{fig:timing-2d} follows the parametric curve $t_c(\Mchirp,\fin) = T_\mathrm{obs}$, with statistics for sources that chirp out-of-band during the observation being less expensive to compute (as the signal occupies fewer tranches).
Other features in \cref{fig:timing-2d} are associated with hardware-specific resource limits that lead to differences in performance as a function of parameter space.
While significantly varying performance over parameter space can limit the efficiency of large batch sizes (as during synchronising operations the GPU is forced to wait for the slowest threads to finish), our tile-based approach to the search ensures that all possible sources within a single parameter space tile have similar $t_c$, so the cost-per-statistic is roughly constant even for large batches of simultaneously evaluated waveforms/search statistics.
Our approach is roughly three orders of magnitude faster than the GPU-accelerated frequency-domain implementation of \cref{eq:Upsilon-seg} employed in previous work (as shown in \cref{tab:search-comparisons}), a sufficient gain in efficiency to enable searches over the full parameter space of mildly eccentric \smBBH{} signals.

\subsection{Implementation details}
\begin{figure*}[t!]
    \centering
    \includegraphics[width=0.99\linewidth]{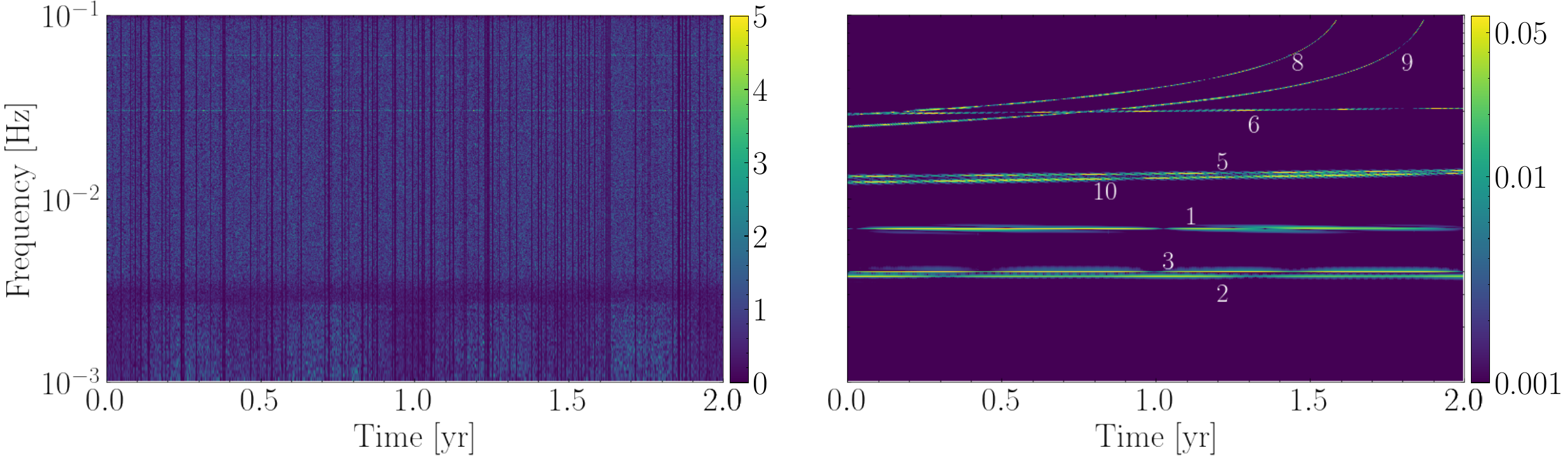}
     \caption{Time-frequency representations of TDI channel $A$ of the \Yorsh{} dataset, containing \smBBHs{}. 
     In both panels, the data are whitened according to the \Sangria{} analytical PSD, which we use in the analysis (see text for more details). 
     Left: The full (GW signals \textit{plus} noise) \Yorsh{} data, with data gaps at 85\% duty cycle. Dark vertical lines correspond to the location of the data gaps. Other features visible by eye are described in the text.
     Right: The GW signal-only contribution to the data (no gaps), showing the time-frequency tracks of the 8 \smBBHs{} present in the data set (labels correspond to the \Yorsh{} source ID, see \ref{tab:injection_params}); all signals are completely invisible in the time-frequency data on the left. Both colorbars quantify the amplitude of the (dimensionless) whitened strain; the left (right) panel is scaled linearly (logarithmically).
     }
    \label{fig:data-time-frequency}
\end{figure*}

Our goal is to detect \smBBHs{} over multiple years of LISA observations. 
The data set consists of the three pseudo-orthogonal TDI time series represented as $d_\alpha(t) = n_\alpha(t) + y_\alpha(t)$, where $\alpha$ labels the second-generation TDI observables $A$, $E$ and $T$~\cite{Prince:2002hp}. 
Here $n_\alpha(t)$ and $y_\alpha(t)$ are the total noise and GW signal contributions to the data stream, respectively. 
In this study, we consider GWs from mildly eccentric (\textit{i.e.} with eccentricity $e \lesssim 0.01$) \smBBHs{} for which the two independent polarisations $h_{+,\times}(t)$ are well-modelled by the dominant $\ell=|m|=2$ mode. We will denote with $\theta$ the set of parameters that describe a GW signal at the LISA output. The TDI observable $y_\alpha(t)$ produced by $h_{+,\times}(t)$ 
can be written as $y_\alpha(t) = \mathfrak{T}_\alpha[t, f, L(t)] h(t)$, where $\mathfrak{T}_\alpha[t, f, L(t)]$ is an operator that projects $h = h_{+} - i h_{\times}$ onto the laser paths connecting the movable optical sub-assemblies of the LISA constellation (the LISA ``arms") with the appropriate time delays. 
The complexity of this response function, which depends on time, $t$, GW frequency, $f$, and the slowly-varying LISA constellation arm-lengths, $L\approx 8.3\,\mathrm{s}$, further complicates  and increases the computational costs of constructing TDI variables, particularly in the frequency band relevant to \smBBHs{} (and EMRIs) where the induced phase shifts, which scale as $\pi L f \approx 0.3 3\,(L/8.3\,\mathrm{sec})\,(f/10\,\mathrm{mHz})$, cannot be ignored. Following Ref.~\cite{Bandopadhyay:2025}, we approximate TDI-2 variables with rescaled TDI-1.5 variables, which account for the rigid rotation of a constant-arm-length constellation \cite{Cornish:2003,2021LRR....24....1T} while ignoring point-ahead corrections (the accuracy of this approximation is quantified in Appendix.~E of Ref.~\cite{Valencia:2025}). The TDI variables are constructed using single-link observables, the transfer functions for these can be found in  Eqs.~(62) and (63) in Ref.~\cite{Marsat:2018}. Overall, our approach is tailored to take maximum advantage of the hierarchy of timescales, $\Tobs \ge \Tseg \ge \Ttr$, that we have introduced at the beginning of this section.

To efficiently explore the detection statistic while exploiting the massively-parallel nature of the problem, we employ a particle swarm optimiser to identify the global maximum of the detection statistic in each parameter tile. For further details of the implementation and settings of this optimiser, see Refs.~\cite{Bandopadhyay:2024,PySO}.


%

\begin{figure*}[t]
    \centering
    \includegraphics[width=0.99\linewidth]{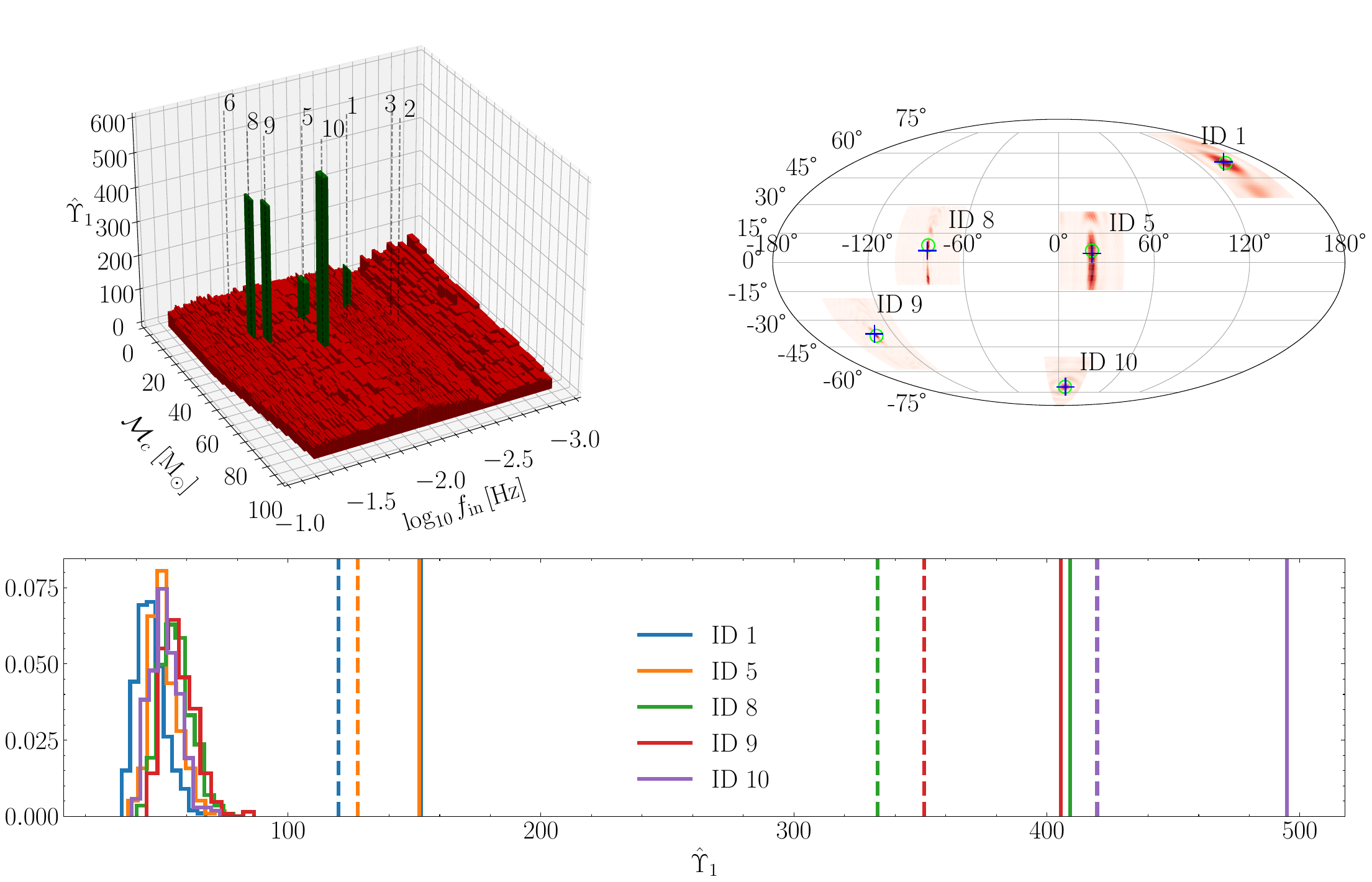}
    \caption{The results of the search on both 100\% and 85\% duty cycle datasets. 
    Top-left: 
    Maximum detection statistic $\hat{\Upsilon}_1 \equiv \mathrm{max}_{\xi}(\Upsilon_1)$
    within each search tile $(\Mchirp, \fin)$. 
    A conservative threshold of $\Upsilon_1\geq 90$ distinguishes significant search triggers (green) from noise triggers (red). 
    Vertical black dotted lines indicate tiles with injected sources. 
    Top-right: Heat-map on the sky of $\Upsilon_{1}(\lambda, \beta)$, conditioned on the search results for each detected source. 
    Blue crosses denote the true location of a source and lime green circles denote the sky position returned by the search. 
    We stress that the density function $\exp(\Upsilon_{1}(\lambda, \beta)/2)$, which could serve as a proposal for estimating the posterior distribution of the source's location in the sky, is significantly more compact than the region covered by the heat-map (see Table 2 in Ref.~\cite{Bandopadhyay:2025}).
    Both top panels refer to the analysis of the 100\% duty cycle dataset.
    Bottom: The distribution of $\hat{\Upsilon}_1$ from background analyses compared to the value of the detection statistic returned by the search for each significant tile, with 
    solid (dashed) lines corresponding to the analysis on the $100\%$ ($85\%$) duty-cycle \Yorsh{} data set. 
    Associated false alarm probability estimates are provided in Table.~\ref{tab:false-alarm}.
    }
    \label{fig:results}
\end{figure*}
%
%

\section{Searching the Yorsh dataset} 
We apply this search pipeline to the most recent LDC release, \Yorsh{} 1.b, see Fig.~\ref{fig:data-time-frequency}. 
This is a $2\,\mathrm{y}$ dataset, sampled at $5\,\mathrm{s}$ cadence, of second-generation TDI variables built under the approximation of constant and unequal constellation arm-lengths.

The instrumental noise is Gaussian and stationary, with an additional cyclo-stationary foreground due to a population of $3\times 10^7$ Galactic ultra-compact binaries with signals of $\mathrm{SNR} > 50$ removed to simulate residuals returned by a global fit analysis of these sources~\cite{yorsh_dataset,yorsh_tech}.
The data also include overlapping signals from $8$ \smBBHs{} with $\Mchirp \approx 7-40\,M_\odot$, mass ratio $q\approx 0.5-1$, effective spin parameter~\cite{Schmidt:2012rh} $|\chi_\mathrm{eff}| \approx 0-0.5$, $\fin \approx 3-30\,\mathrm{mHz}$, where $\fin \equiv f(t_0)$, (corresponding to  
$t_c \approx 1.5 - 300\,\mathrm{yr}$) generated using the
spin-aligned and quasi-circular \texttt{IMRPhenomD} waveform approximant~\cite{Khan:2016}. Injection optimal SNRs $\rho^\mathrm{inj}$ range from $\approx 8-25$, with parameters given in \cref{tab:yorsh1b_injections}.

In the search, we employ the waveform approximant \texttt{TaylorF2Ecc} which allows us to search for spin-aligned and eccentric binaries \cite{Moore:2016qxz,2022PhRvD.105b3003F}, but in the search templates we set $\chi_\mathrm{eff} = 0$; as was identified in Ref.~\cite{Bandopadhyay:2025}, aligned spin effects introduce small deviations to the inspiral phase that can be absorbed by a phase maximisation and small parameter shift with little to no loss in SNR, and we find the inclusion of $\chi_\mathrm{eff}$ does not impact search performance. This parameter can be readily determined in \textit{e.g} a follow-up Bayesian parameter estimation analysis, which we do not perform in this work, but has been demonstrated previously in Refs.~\cite{Bandopadhyay:2025,Bandopadhyay:2024}. 
The search parameter vector on which the detection statistic, Eq.~(\ref{eq:Upsilon}), depends is $\xi = \{\Mchirp, q, \fin, \ein, \iota, \psi, \lambda, \beta\}$, where $\ein\equiv e(t_0)$, $\psi$ is the polarisation angle, $\iota$ the inclination angle of the orbital plane, and $(\lambda, \beta)$ identify the source location in the sky through the ecliptic longitude and latitude, respectively. The search covers $\Mchirp \in [5, 100]\,M_\odot$, $q \in [0.1, 1]$, $\ein \in [0, 0.01]$ and $\fin \in [1, 100]\,\mathrm{mHz}$, exploring the entire sky and all inclinations and polarisation angles.

In constructing the time-frequency representation of $y_\alpha(\xi)$, we set a tranche duration of $\Delta T = 6\, \mathrm{hr}\,(3\,\mathrm{hr})$ for $\fin$ below (above) $10\, \mathrm{mHz}$.
Shorter tranches are used for higher $\fin$ in order to better retain the accuracy of the linear approximation to the frequency evolution.
Each tranche is Tukey-windowed with a shape parameter of $0.1$ and fast-Fourier-transformed to construct the time-frequency grid over which our analysis is conducted.
We model the noise PSD in each tranche according to the \textit{Sangria} LDC model~\cite{ldcsoft}, which accounts for both instrumental and Galactic foreground noise sources~\cite{sangria_tech}.
This PSD crucially does not model the cyclo-stationarity of the Galactic foreground, leading to the periodic variations below $\sim 3\,\mathrm{mHz}$ in the left panel of~\cref{fig:data-time-frequency}.
Note the dip in the whitened spectrogram at $\sim 3\,\mathrm{mHz}$: the foreground contribution to the \textit{Sangria} PSD overestimates confusion noise in this region, which manifests as a feature in our results (\cref{fig:results}) at low frequencies.
The spurious horizontal lines in the left panel of Fig.~\ref{fig:data-time-frequency} are caused by inconsistencies between the \Yorsh{} data and the \textit{Sangria} PSD. Despite these inconsistencies, our pipeline is sufficiently robust to noise modelling errors that the results of our search are not strongly impacted.

To ensure even coverage of the parameter space and enable parallelisation over a pool of GPUs, we tile the search volume in $(\Mchirp, \fin)$.
Each tile has dimensions of $\Delta \Mchirp=5\,\mathrm{M}_\odot$ and $\Delta \fin = $$10^{-4}\,(10^{-3})\,\mathrm{Hz}$ for $\fin$ below (above) $10^{-2}\,\mathrm{Hz}$, chosen to approximately balance the per-tile computational cost (the GPU wall-time per tile is $\lesssim 0.5\,\mathrm{h}$, see \cref{tab:search-comparisons}).
We exclude a small number of tiles for the largest values of $\Mchirp$ and $\fin$, as they correspond to signals for which $t_c \ll \Tobs$; in this regime the slow GW frequency evolution assumption central to the efficient implementation of the search breaks down. We then search over each of the remaining $2680$ tiles with 4 NVIDIA A100 GPUs, which is completed in $\sim 11$ days. Results are shown in \cref{fig:results}. 

\subsection{Search results on the full dataset}

In the upper-left panel of~\cref{fig:results}, we show the maximum of the detection statistic, $\hat{\Upsilon}_1 \equiv \mathrm{max}_{\xi}(\Upsilon_1)$, returned by the search within each tile. We consider a tile to be statistically significant if $\hat{\Upsilon}_1$ crosses an (arbitrary) threshold of $\Upsilon_1=90$, corresponding to a false alarm probability which differs from tile to tile but is $\sim 10^{-4}$. 
To quantify the detection significance of each source candidate, we construct $\hat{\Upsilon}_1$ background distributions by repeating an identical search over $300$ realisations of noise-only data (as described by our noise model) for tiles in which $\Upsilon_1 > 90$, see bottom panel of~\cref{fig:results}. 
The returned $\hat{\Upsilon}_1$ lie far outwith the background distributions in all cases, and we estimate the false alarm probability of these signals to be $\ll 10^{-5}$. 
Note that the background distributions for the 85\% duty cycle data set (whose analysis we describe below) are statistically equivalent to those of the full \Yorsh{} dataset (as they are ultimately only a function of $\Nseg$ and the dimensions of the search space).

We further verify these detections by examining $\Upsilon_1$ as a function of sky position conditioned on all other parameters (right panel of \cref{fig:results}), finding that the location of each injected source agrees precisely with the bulk of $\Upsilon_1$ returned by the search. 
In~\cref{sec:results}, we also show that the values of $(\Mchirp, \fin)$ returned by the search are consistent to better than one part in $\sim 10^4$ with the injection parameters for each of these sources.

The search detects all signals with $\rho^\mathrm{inj} \gtrsim 10$, with the exception of source 6. Notably, source 1 $(\rho^\mathrm{inj} \approx 11)$ is correctly identified, which would not have been possible if our segment ladder started at $\Nseg=100$~\cite{Bandopadhyay:2025}.

After exploring this result in more detail, we confirmed that this is because source 6 occupies a smaller fraction of the search volume than other signals of lower SNR (likely due to its low chirp mass), see further discussion in \cref{sec:Source_6}.
This is an artefact of our relatively simple approach for tiling the $(\Mchirp, \fin)$ space; we expect that future implementations using tiles with size chosen according to the region of parameter space will be more robust to this behaviour.

\subsection{Search extension to data with gaps}
The time-frequency framework offers significant flexibility when considering signals of duration longer than the typical time-scales over which noise is stationary and data collection suffers interruptions.
For LISA, galactic confusion noise (which is expected to dominate instrumental noise below $\approx 3\,\mathrm{mHz}$) is cyclo-stationary and changes on a time-scale of $6$ months, and the instrumental noise itself is expected to vary on timescales $\ll \Tobs$.
Moreover, gaps (either due to scheduled instrument adjustments or unforeseen data dropouts or contaminations~\cite{Baghi:2021tfd}) will be present; the LISA science data duty cycle requirement is $> 82\%$~\cite{2024arXiv240207571C}. 
Viable strategies for LISA analysis must be robust to such features of the data stream.

\begin{table*}[htb!]
    \caption{ \label{tab:injection_params}
    Source parameters of the \smBBHs{} present in the \Yorsh{} LDC dataset.
    The final column reports the optimal SNR of each injection ($\rho^{\rm inj}$) computed using the injected waveforms (from the noiseless data streams) and the LDC \Sangria{} PSD. We (arbitrarily) set the zero reference time to the beginning of the LISA data, hence the time of coalescence, $t_c^{\rm inj}$, also corresponds to remaining binary lifetime from the start of LISA observations.
    All of the injected binaries were on quasi-circular orbits (zero eccentricity). 
    Sources marked with a star, ``*", are those that are identified by our search pipeline.} 
\begin{ruledtabular}
\begin{tabular}{|l|cccccc|c|}
    \textrm{ID} &
    $\mathcal{M}_c^{\rm inj} \,  [\mathrm{M}_{\odot}]$ &
    $t_c^{\rm inj} \, [\rm{y}]$ & 
    $\fin^{\rm inj}\, [\rm{mHz}]$ & 
    $D_L^{\rm inj} \, [\rm{Mpc}]$ & $q^{\rm inj}$ & $\chi_{\rm eff}^{\rm inj}$ &
    $\rho^{\rm inj}$\\
    \hline \hline 
    1* & $29.34741587$ & $\phantom{0}65.9176$ & $\phantom{0}5.85830665$ & $159.9$ & $0.91$ & $\phantom{-}0.50\phantom{0}$& $10.91$ \\
    2 & $40.3876144$ & $281.3679$ & $\phantom{0}2.78485696$ & $\phantom{0}93.06$ &  $0.83$ & $ -0.23 $ & $8.28$ \\
    3 & $34.51216704$ & $297.7712$ & $\phantom{0}3.00698596$ & $\phantom{0}47.0$ &  $0.58$ & $\phantom{-}0.10\phantom{0}$ & $9.88$ \\
    5* & $27.41970433$ & $\phantom{0}10.3457$ & $12.24273032$ & $168.3$ &  $0.83$ & $-0.55\phantom{0}$ & $12.93$ \\
    6 & $7.007404972$ & $\phantom{0}11.0420$ & $28.02352272$ & $\phantom{0}17.3$ &  $0.88$ & $-0.17\phantom{0}$ & $13.99$ \\
    8* & $22.40969304$ & $\phantom{00}1.6501$ & $27.65438527$ & $\phantom{0}34.0$ &  $0.59$ & $\phantom{-}0.002$ & $23.74$ \\
    9* & $26.08583360$ & $\phantom{00}1.9185$ & $23.76783772$ & $\phantom{0}85.5$ &  $0.95$ & $\phantom{-}0.10\phantom{0}$ & $23.94$\\
    10* & $39.14942200$ & $\phantom{00}7.0604$ & $11.31112717$ & $168.9$ &  $0.88$ & $\phantom{-}0.03\phantom{0}$ & $24.66$ 
\end{tabular}
\end{ruledtabular}
\label{tab:yorsh1b_injections}
\end{table*}

To examine the performance of our approach under such conditions, we zero-out random portions of the \Yorsh{} data stream to simulate the presence of gaps and produce an $85\%$ duty-cycle \Yorsh{} dataset. 
Gaps are produced at random times with 6 hour durations, accumulating a total data dropout of $15\%$ over the observation period.
The spectrogram of the resulting data is shown in the left panel of \cref{fig:data-time-frequency}.
We then re-process the data with the same pipeline, with the only change being that tranches that overlap with data gaps are not included in the computation of the time-frequency inner-product~[Eq.~(\ref{eq:inner-tf})], and therefore do not contribute to the detection statistic~[\cref{eq:Upsilon}]. 
For further flexibility, our framework permits a time-frequency PSD that varies between \textit{tranches}, which can be used to model noise spectra which are locally stationary~\cite{Digman:2023,2022ApJ...940...10D}; we do not demonstrate this capability in this work, but have verified this performs as expected for our implementation.

The end-to-end data processing remains efficient and is consistent with the results shown in~\cref{tab:search-comparisons}. 
The results of this analysis are shown in~\cref{fig:results}, and additional results are presented in~\cref{sec:results}. 
The bottom panel of~\cref{fig:results} shows that all of the sources that were identified in the full \Yorsh{} data set are still recovered when gaps are present.
As expected, $\Upsilon_1$ is smaller in all cases, as $\Upsilon_1 \approx \rho^2$, 
which has decreased due to the loss of data.
The fact that the values of $\Upsilon_1$ obtained are $\approx 15\%$ less than in the search of the full \Yorsh{} data (in close agreement with the loss in $\rho^2$ expected for a perfectly monochromatic signal) verifies that our approach is capable of analysing data containing frequent gaps without incurring additional sensitivity losses or sacrificing  computational performance.

\subsection{Injected and recovered signals}\label{sec:results}

Here we provide additional details of the signals injected in the LDC data set \Yorsh, as well as results supporting the accuracy and significance of the detections reported by our search pipeline. 

\Cref{tab:injection_params} shows the details of the binaries whose GW signal was included in the \Yorsh{} LDC dataset. 
The \Yorsh{} catalogue does not provide $t_c$; the values provided in \cref{tab:injection_params} are inferred from the provided parameters using the \texttt{TaylorF2Ecc} model (with $e=0$). 
We also report the optimal signal-to-noise ratio of the injections, computed with the LDC \Sangria{} PSD model used in the search.
Our source naming convention follows that of the \Yorsh{} technical document~\cite{yorsh_tech}; note that source IDs 4 and 7 are not included in the table because these signals are not present in the data set.
\begin{table}[htb!]
    \caption{
    False alarm probabilities extrapolated from $\Upsilon_1$ background distributions in search tiles passing the detection threshold. 
    The power-law and Gumbel fits provide conservative and optimistic estimates respectively (see text for details).
    }
    
\begin{ruledtabular}
\begin{tabular}{|cc|cc|cc|}
    \hline
     \multicolumn{2}{|c|}{Source} & \multicolumn{2}{c|}{Power law} & \multicolumn{2}{c|}{Gumbel} \\
    \cline{3-6}
    \textrm{ID} & $\rho^\mathrm{inj}$            & 100\% & 85\%  & 100\% & 85\% \\
    \hline \hline 
    1 & 10.91 & $5.6 \times 10^{-7}$ & $8.7 \times 10^{-6}$ & $9.2 \times 10^{-11}$  & $9.0 \times 10^{-8}$ \\
    5 & 12.93 & $3.2 \times 10^{-8}$ & $4.9\times 10^{-7}$ & $8.8 \times 10^{-10}$ & $1.2\times 10^{-7}$ \\
    8 & 23.74 & $5.2 \times 10^{-10}$ & $4.4 \times 10^{-9}$ & $\sim 10^{-28}$ & $\sim 10^{-22}$ \\
    9 & 23.94 & $3.5 \times 10^{-13}$ & $2.7 \times 10^{-12}$ & $\sim 10^{-28}$ & $\sim 10^{-24}$ \\
    10 &  24.66 & $8.3 \times 10^{-18}$ & $1.5 \times 10^{-16}$ & $\sim 10^{-39}$ & $\sim 10^{-32}$ \\
    \hline
\end{tabular}
\end{ruledtabular}
    \label{tab:false-alarm}
\end{table}

Assessing statistical significance of candidates will generally be non-trivial for LISA data due to the challenge of generating (many statistically identical) signal-free data sets. Here, we calculate the false alarm probabilities for source candidates by comparing against a background distribution computed using \textit{simulated} noise-only data; this quantity corresponds to the probability that the search will find a source candidate with $\Upsilon_1\geq \Upsilon_{1|\rm{candidate}}$. 
This involves generating several statistically independent realisations of noise-only data and running the search on each realisation, empirically building the background distribution of $p_0(\Upsilon_{1})$. Following Ref.~\cite{Bandopadhyay:2024}, we extrapolate this distribution using a power-law tail (conservative) or Gumbel (optimistic); these extrapolated fits are used to compute the upper and lower bounds for the false-alarm-probabilities for each of the candidate detections. Note that the background distributions in Fig.~\ref{fig:results} are computed individually for each search tile (where there is a candidate detection); the variation in the background distributions is caused by the difference in the prior-to-posterior ratio across parameter space; using a metric-based approach (such as that employed in Ref.~\cite{Harry:2009}) to place search tiles would reduce the differences between these distributions. 

In \cref{tab:false-alarm}, we report false alarm probabilities with which the signals are recovered in the search, for both the full ($100\%$) and $85\%$ duty cycle \Yorsh{} datasets. In the case of a power-law extrapolation, we fit $\delta$ for $p_0(\Upsilon_1) \propto \Upsilon_1^{\delta}$; for the Gumbel distribution $p_0(\Upsilon_1) \propto G(\mu;\beta)$, we fit the free parameters $\mu$ and $\beta$ (location and scale, respectively).

 \begin{figure}[htb!]
    \centering
    \includegraphics[width=0.9\linewidth]{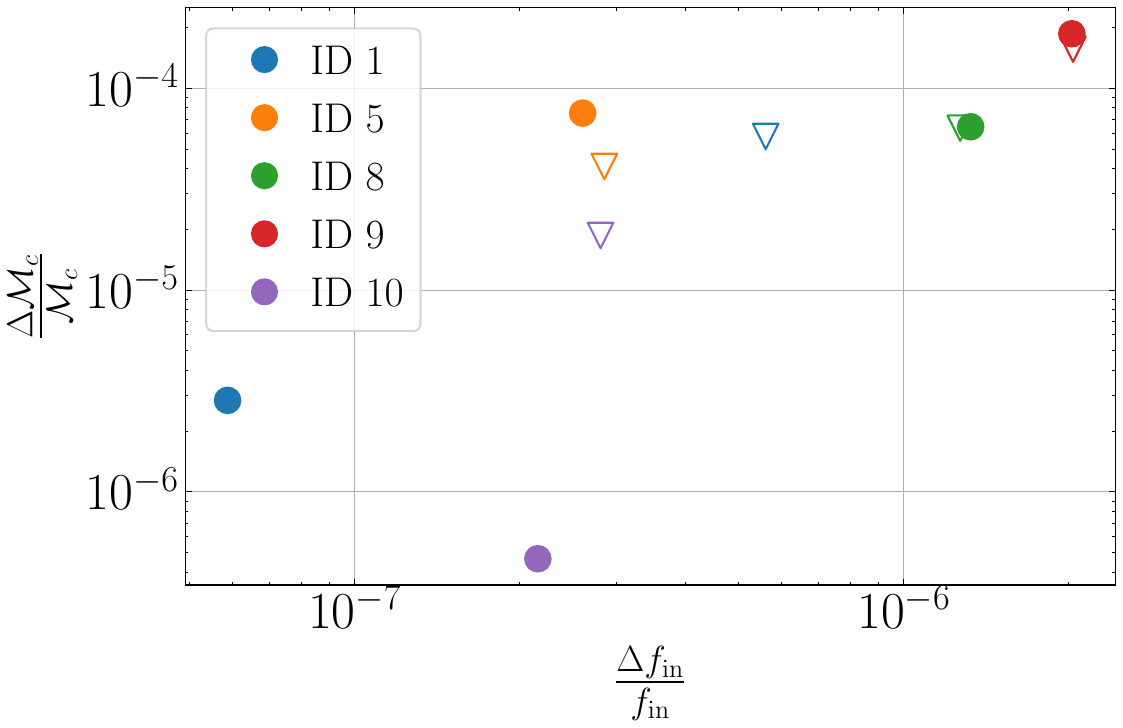}
    \caption{
    Relative error in the recovery of $\Mchirp$ and $\fin$ (with respect to injected values) for detected signals at either $100\%$ (filled circles) or $85\%$ (triangles) duty cycle.
    As we search with a stochastic optimiser, some variability in the maximum-statistic parameters is expected, which can cause relative error to decrease despite the loss in SNR when gaps are present.
    }
    \label{fig:fract_errors}
\end{figure}

To demonstrate that our search pipeline successfully \textit{identifies} \smBBHs{}, as opposed to merely detecting them, we also compute relative errors in the recovered $\Mchirp$ and $\fin$ for each source relative to their injected values (\Cref{tab:injection_params}), which complements the results concerning the sky location that are reported in Fig.~\ref{fig:results}.
The results are shown in~\cref{fig:fract_errors}. In all cases, the injected and recovered parameters are in close agreement, with relative errors on scales typical of the measurement precision of these parameters, see \textit{e.g.} Ref.~\cite{Buscicchio:2021, Bandopadhyay:2025}. The results of the analysis can then be used to seed a dedicated follow-up parameter estimation stage, which can efficiently return posterior distribution functions on the source parameters.

\subsection{Analysis of source ID 6}\label{sec:Source_6}
Source 6 is the only injected signal with $\rho>10$ that was not found by the search. 

Our search can fail to identify a signal for one of two reasons: (i) $\rho$ is too low for the maximum segment count of $\Nseg=50$ (\textit{i.e.}, the signal's statistic value is not significant with respect to noise fluctuations), or (ii) the peak in the search surface is too narrow at $\Nseg=50$ to be reliably encountered by the stochastic optimiser.
While sources 2 and 3 are not identified due to the former reason (as their SNRs are lower, $\rho^{\rm inj} \approx 8-9$), source 6 with $\rho^{\rm inj} \approx 14$ is not found due to the latter reason. Indeed, we have tested that the signal is reliably detected by our pipeline once the search prior is narrowed by a factor of $\sim 4$.

\begin{figure*}
    \centering
    \includegraphics[width=0.99\linewidth]{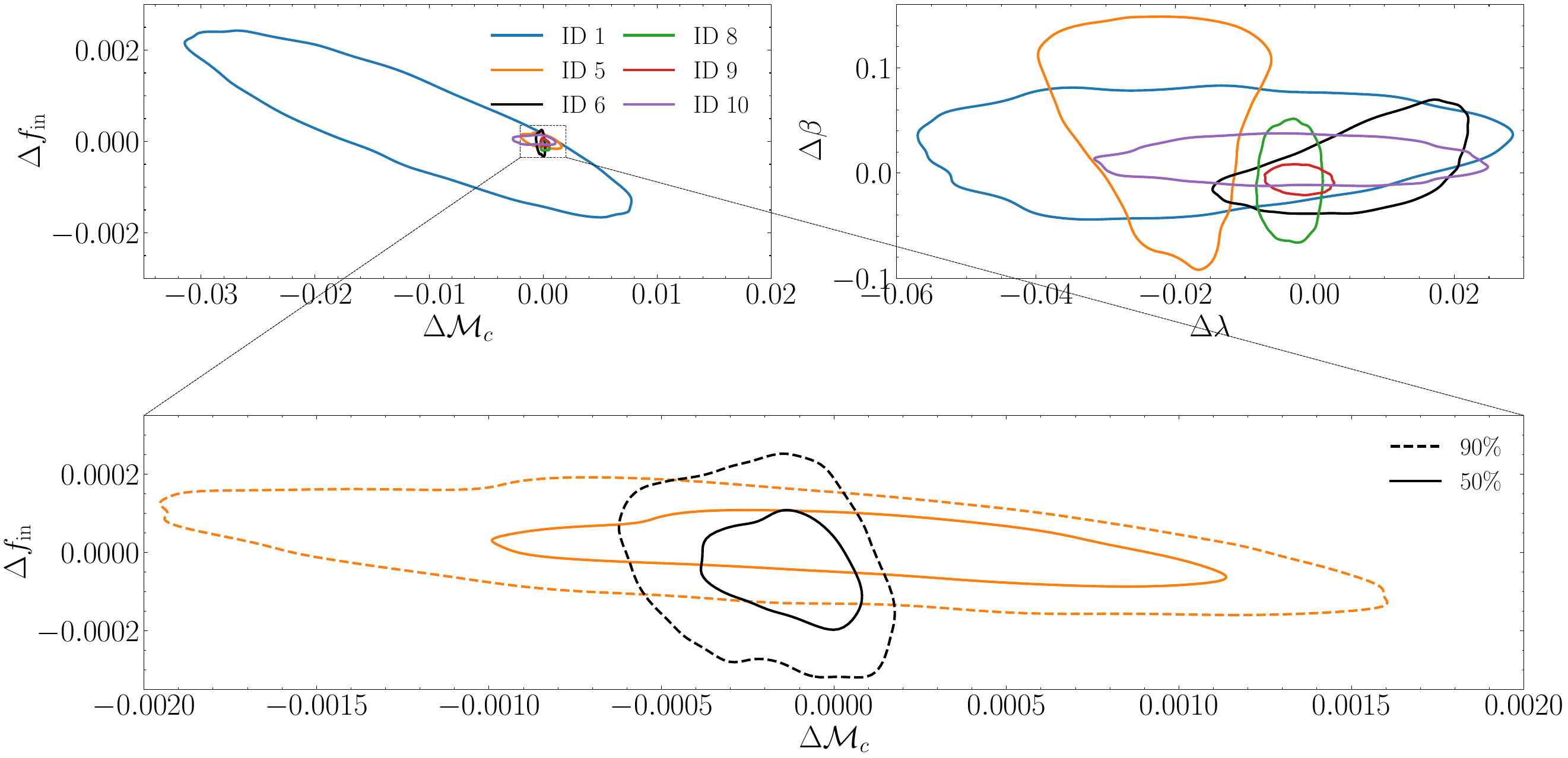}
    \caption{Comparison of marginal posterior distributions for source 6 (not detected by our pipeline) with successfully identified sources in the \Yorsh{} dataset (full duty cycle). 
    Top panel: Marginal posterior distributions of $p(\xi) \propto \exp(\Upsilon_{50}/2)$ in $(\Mchirp, \fin)$ and $(\lambda, \beta)$, presented in units of search space dimensions, for the \Yorsh{} injections (\cref{tab:injection_params}).
    Bottom panel: Closer view of the $(50\%, 90\%)$ intervals for sources 5 and 6, which are similar in SNR and $\fin$ (and thus are the most comparable).
    The larger extent of the distribution for source 5 (particularly with respect to $\Mchirp$) explains why this source was identified while source 6 was not. 
    We emphasise that these are \textit{not} coherent marginal posterior distributions. The nested sampling implementation within \texttt{nessai} was used to stochastically sample these distributions \cite{nessai,Williams:2021qyt}.}
    \label{fig:Overlay_PE_N_50}
\end{figure*}

To understand this behaviour, we performed targeted exploration of the primary mode of the posterior distributions $p(\xi) \propto \exp(\Upsilon_{50}/2)$ for each source via nested sampling, as shown in~\cref{fig:Overlay_PE_N_50}. 
This confirms that source 6 occupies a smaller fraction of the prior space than source 5 (the most comparable source in both SNR, $\rho^{\rm inj} \approx 13$, and $f_{\rm{in}}$), which explains why this source was identified despite having a lower SNR.
While the $50\%$ confidence intervals of the posterior distributions for sources 8 and 9 (which were both successfully identified) occupy a smaller region of parameter space than source 6, the tails of these distributions extend much further across parameter space before sinking into the noise floor due to their larger SNR.
This means that the particle swarms are far more likely to encounter (and then climb) the deformations in the statistic manifold corresponding to these sources.
The volume occupied by the primary mode (for constant SNR) is an intrinsic property of the waveform model (and hence the search statistic) as a function of the $(\Mchirp,\fin)$ parameter space.
An adaptive tiling strategy that accounts for this behaviour (similarly to what is employed in the construction of template banks for ground-based searches~\cite{Owen:1995tm,Sakon:2022ibh}) and redistributes search efforts will improve the reliability of the search at little additional cost. 
It is our expectation that such a tiling strategy will allow us to find source 6. 
We will revisit strategies for parameter-space tiling in future work.

\section{Outlook} 
The time-frequency GPU-accelerated search approach for long-duration, broad-band GW signals that we have successfully demonstrated here provides a solution to the long-standing problem of searching globally for \smBBHs{} in LISA. 
In the most na\"ive version in which each tile is assigned to a GPU, it meets the requirement of completing a global search every few weeks as the mission progresses -- the $\mathrm{SNR}$ accumulates slowly, approximately $\propto \Tobs^{1/2}$ --  while achieving $\lesssim 1\,\mathrm{hr}$ latency in specific tiles, should detection candidates, and the need for alerts, start to emerge. 
Data processing in $\approx \mathrm{hr}-$long time segments seamlessly accounts for data-gaps and the complexity of noise non-stationarities over the (much longer) observation timescale \cite{2025arXiv251006406B,2025PhRvD.111l4053B}. 

There are several extensions and improvements to the approach presented here that we intend to pursue in the future. While in this study analytic orbits are used to model the spacecraft \cite{Cornish:2002rt}, this can be generalised, at no additional computational costs to numerical and fully generic orbits with varying arm-length and point-ahead corrections \cite{2021JAnSc..68..402M,PhysRevD.106.103001}. A more rigorous metric-based approach to the $(\Mchirp, \fin)$ tiling would ensure that the ratio between the search volume vs the posterior volume is constant across the parameter space, solving the issue that we have encountered with source 6 in \Yorsh~\cite{Owen:1995tm,Sakon:2022ibh}. 
We plan to further expand the search parameter space to cover black holes with arbitrary spin orientation (which induces precession of the orbital plane and modulations of the emitted GWs), orbits with large eccentricity and radiation's multiple modes, all of which will be particularly pertinent when tackling the even more challenging problem of EMRI identification. The time-frequency representation of the waveform can be extended to the multi-harmonic case, simply by computing the waveform quantities  $\{A,f, \dot{f},\phi\}$ for each harmonic, and subsequently applying the Fresnel expressions to each harmonic.

The increased computational burden incurred by considering waveforms which include additional physical effects discussed above can be mitigated by using a parameter space tiling scheme (e.g. as done here over $(f_0 ,\mathcal{M}_c)$) which is \textit{adaptive}. Such a tiling scheme would have greater resolution (i.e. more tiles) in parts of parameter space where the posterior is expected to be narrower. The search is carried out independently in each tile, and as a result the cost of the search can be crudely estimated to scale as $\mathcal{O}(N_{\rm{tiles}}/N_{\rm{GPUs}})$. Additionally, improvements to GPU hardware, particularly those which allow use of higher numbers of concurrent threads, will also significantly speed up the methods presented here.

More broadly, our approach is applicable to a wide variety of GW searches, both for LISA and the current ground-based detector network, providing a viable path to expanding the parameter space that can be covered.

\section*{Data availability} 
Data supporting the Figures, Tables and conclusions of this work are openly available~\cite{christian_chapman_bird_2026_21835669}.

\section*{Acknowledgements} 
D.B. is supported by MPG and acknowledges support from UK Space Agency grant UKRI971. C.E.A.C-B. and A.V. are supported by UKSA Space Agency grant UKRI971. A.V. acknowledges the support of the Royal Society and Wolfson Foundation. 
A.V. thanks the Flatiron Institute for hospitality while a portion of this research was carried out. 
The Flatiron Institute is a division of the Simons Foundation. 
The computations for this paper were performed using the University of Birmingham's BlueBEAR High Performance Computing facility.


\bibliographystyle{apsrev4-2}
\bibliography{bibliography}


\clearpage
\appendix
\begin{figure}[htb!]
    \centering
    \includegraphics[width=0.99\linewidth]{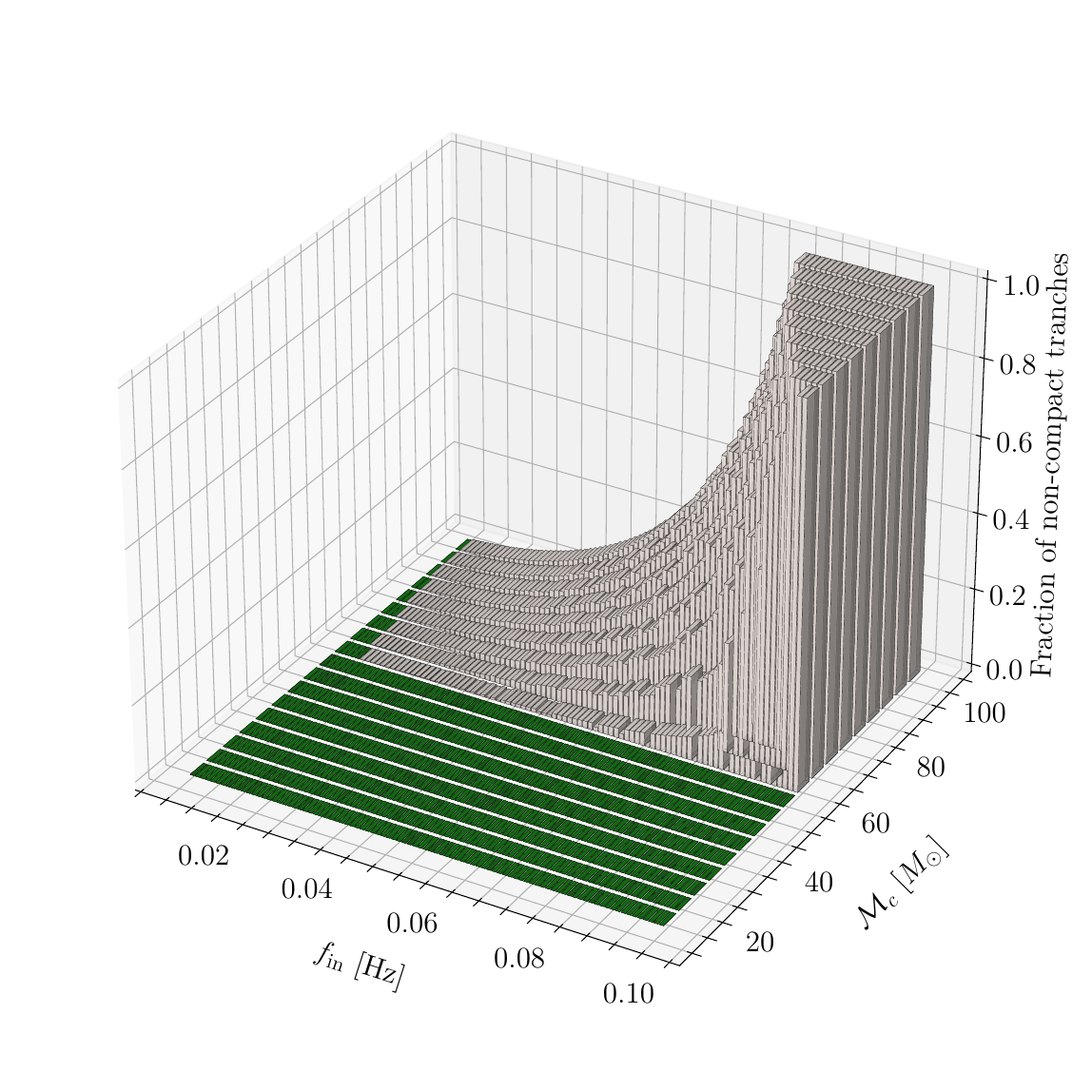}
    \caption{Fraction of SFT tranches which violate the `compactness condition', for simplicity we consider the signal to be compact within a tranche if it satisfies the condition $\dot{f}\Delta T \leq 10 \delta f$, i.e. it crosses less than $11$ Fourier bins within $\Delta T$. A signal violating the compactness condition does not necessarily imply that the Fresnel integral is no longer valid within that tranche, as shown in the bottom row of Fig.~\ref{fig:validity_of_fresnel_num_bins}.
    }
    \label{fig:invalid-tranches}
\end{figure}
\section{Compactness of the Fresnel-kernel function}\label{app:compactness}

\begin{figure*}[b]
    \centering
    \includegraphics[width=0.99\linewidth]{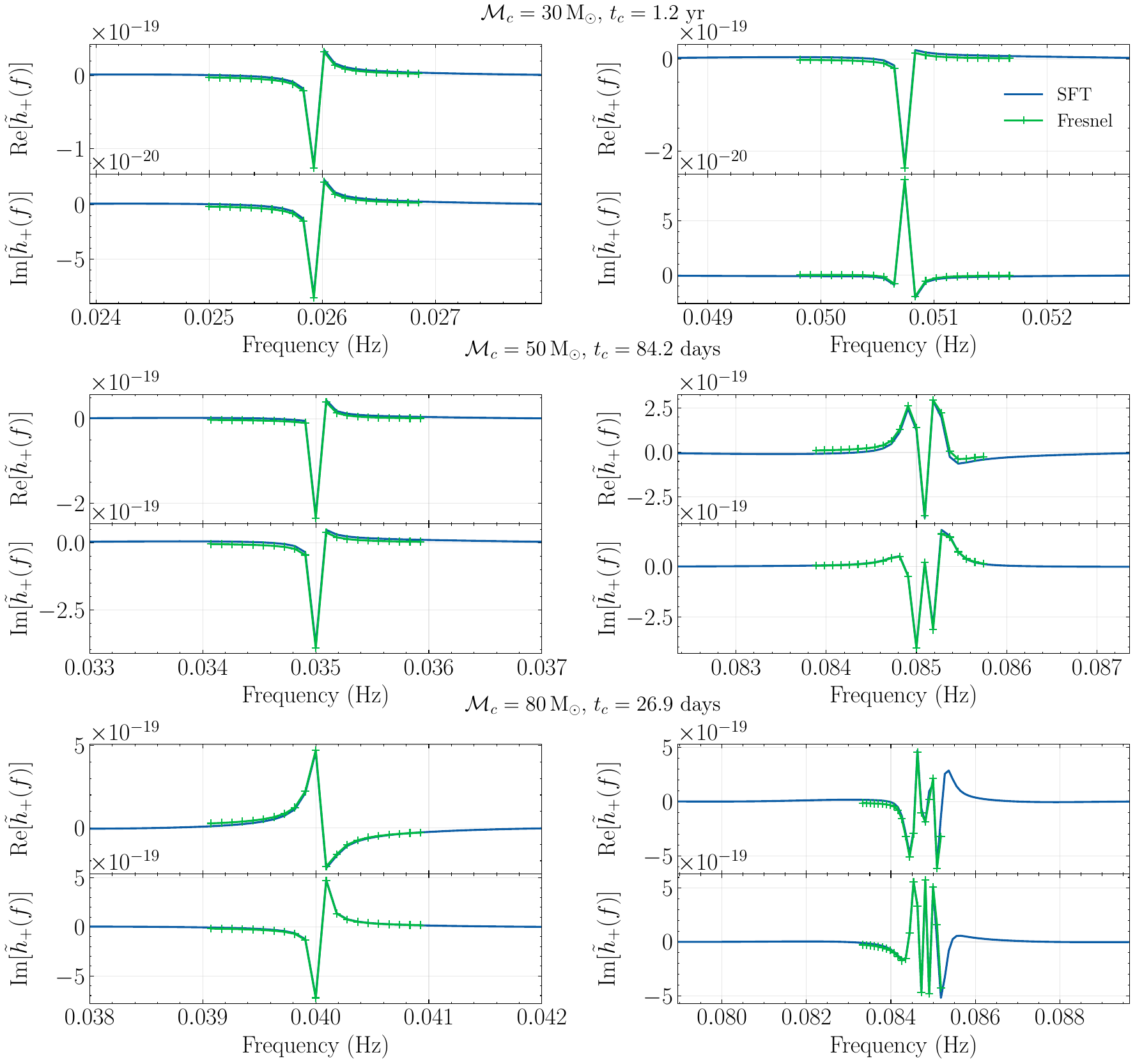}
    \caption{
    Fresnel-kernel approximation of the Fourier transform of a waveform.  Waveforms generated using the Fresnel-kernel function over a tranche (green lines) are compared to the short-Fourier transformed (SFT) time domain waveform (blue lines) in the first (left panels) and last (right panels) SFT time-interval. Top panel: a mildly chirping source ($\Mchirp = 30\, \rm{M}_\odot$) that remains in the LISA frequency band over the whole observation period (in this figure $T_{\rm{obs}} = 1\, \rm{yr}$). Middle panel: a moderately chirping source ($\Mchirp = 50\, \rm{M}_\odot$) that exits the LISA band after 80 days. Bottom panel: a highly chirping source ($\Mchirp = 80\, \rm{M}_\odot$) that exits the LISA frequency band within less than a month. Waveforms computed using the Fresnel-kernel function are generated at 20 points around $f_{nm}$, which is sufficient for the full evolution of the mildly and moderately chirping sources. However, in the last tranche of the highly chirping source, the SFT of the waveforms is not captured fully by the approximation (bottom right plot). This illustrates that the compactness of the Fresnel-kernel breaks down in the late inspiral for highly chirping systems. Note that in this case extending the number of frequency points over which the Fresnel-kernel function is computed would recover the SFT waveform, but add to the computational burden. For simplicity, in this figure we have computed waveforms retaining only the zero-th order post-Newtonian term of the waveform phase, which is fully described by the chirp mass and time-to-merger.}
    \label{fig:validity_of_fresnel_num_bins}
\end{figure*}

The compactness of the Fresnel-kernel function, \cref{eq:kernel-function}, \textit{i.e} the fact that its evaluation over 21 frequency bins centred on $f_{nm}$ for each tranche provides a faithful approximation of the Fourier transform of a waveform via \cref{eq:fresnel-ft}, is one of the key elements that enables the orders of magnitude reduction in computational cost demonstrated in this papers. Here we discuss the variation of this compactness across the parameter space of the search that we have presented. The Fresnel-kernel function remains compact over the vast majority of the $\Mchirp \in [5, 100]\,M_\odot$ and $\fin \in [1, 100]\,\mathrm{mHz}$ parameter space, with the exception of a small number of tiles at the edge of the search space $(f_{\rm{in}}\gtrsim 0.05 \, \rm{Hz}, \mathcal{M}_c \gtrsim 60 \, \mathrm{M}_{\odot})$. However, the systems in this region are in band for $\lesssim 1$ month, and as a result are unlikely to accumulate enough SNR to be detectable in LISA as they would need to be unreasonably close. Additionally, the loss of compactness is limited to the tranches containing the portion of a signal that is about to exit the LISA frequency band, at which point the noise PSD is rising steeply and thus the contribution to the inner product is strongly suppressed from this tranche.  The parameter space dependence of this compactness is illustrated in \cref{fig:invalid-tranches}, the impact of this behaviour on the evaluation of the Fourier transform of a signal using \cref{eq:fresnel-ft} is shown for selected values of $f_{\rm{in}}$ and $\mathcal{M}_c$ in Fig.~\ref{fig:validity_of_fresnel_num_bins}.

\end{document}